\documentclass[]{interact}
\usepackage[colorlinks=true,breaklinks=true]{hyperref}
\usepackage{subfigure}
\usepackage{epstopdf}

\usepackage[longnamesfirst,sort]{natbib}
\bibpunct[, ]{(}{)}{;}{a}{,}{,}

\theoremstyle{plain}

\theoremstyle{definition}

\theoremstyle{remark}

\newcommand{\be}{\begin{equation}}
\newcommand{\en}{\end{equation}}

\def\d{{\rm d}}
\def\uv{{\boldsymbol u}}

\def\gv{{\boldsymbol g}}

\def\rv{{\boldsymbol r}}

\def\kv{{\boldsymbol k}}

\def\ev{{\boldsymbol e}}
\def\Bv{{\boldsymbol B}}

\def\Lv{{\boldsymbol L}}

\def\grad{\boldsymbol\grad}

\def\grad{{\rm grad}\, }
\def\bnab{{{\boldsymbol\nabla}}}
\def\bdot{{\boldsymbol\cdot}}

\usepackage{epstopdf}
\usepackage{xcolor,natbib}

\newcommand{\dd}{\partial}

\begin{document}

\jvol{00} \jnum{00} \jyear{2012} 



\title{A study of global magnetic helicity in self-consistent spherical dynamos}

\author{P.~Gupta$^{\ast}$$^1$\thanks{$^1$P.~Gupta:~\href{https://orcid.org/0000-0002-2976-5993}{orcid.org/0000-0002-2976-5993}}, R.D.~Simitev$^{\ast}$${^2}$\thanks{$^2$R.D.~Simitev:~\href{https://orcid.org/0000-0002-2207-5789}{orcid.org/0000-0002-2207-5789}} and  D.~MacTaggart$^{\ast}$${^3}$\thanks{$^3$D.~MacTaggart:~\href{https://orcid.org/0000-0003-2297-9312}{orcid.org/0000-0003-2297-9312}}$^{\dag}$\thanks{$^\dag$Corresponding author. Email: david.mactaggart@glasgow.ac.uk
\vspace{6pt}}\\\vspace{6pt}  ${^\ast}$School of Mathematics and Statistics, University of Glasgow, Glasgow, G12 8QQ, UK\\ 
}
\maketitle

\begin{abstract}
Magnetic helicity is a fundamental constraint in both ideal and resistive magnetohydrodynamics. Measurements of magnetic helicity density on the Sun and other stars are used to interpret the internal behaviour of the dynamo generating the global magnetic field. In this note, we study the behaviour of the global relative magnetic helicity in three self-consistent spherical dynamo solutions of increasing complexity. Magnetic helicity describes the global linkage of the poloidal and toroidal magnetic fields (weighted by magnetic flux),  and our results indicate that there are preferred states of this linkage. This leads us to propose that global magnetic reversals are, perhaps, a means of preserving this linkage, since, when only one of the poloidal or toroidal fields reverses, the preferred state of linkage is lost. It is shown that magnetic helicity indicates the onset of reversals and that this signature may be observed at the outer surface. 


\end{abstract}

\section{Introduction}

Magnetic helicity is an invariant of ideal magnetohydrodynamics (MHD) that describes field line topology weighted by magnetic flux \citep{moffatt1969degree,bergerfield1984}. This quantity is important for both laboratory \citep{taylor1986relaxation} and astrophysical plasmas \citep{petsov2014,mactaggart2020} since it remains approximately conserved in the weakly resistive limit \citep{bergerRigLim1984,faraco2019,faraco2022}. Magnetic helicity has been studied, in detail, in the context of mean field dynamo models \citep[see][and references within]{brandenburg2005astrophysical,cameron2017global,brandenburg2018}. In self-consistent convection-driven dynamos in spherical shells, however, magnetic helicity has received little attention. On the scales typically considered in such global simulations, it is not possible to reach the high magnetic Reynolds number range needed to maintain an approximately constant value of the global magnetic helicity. This does not mean, however, that magnetic helicity is not a useful measure for understanding the nature of self-consistent spherical dynamos. In this work, rather than considering the role that magnetic helicity plays in turbulence (such as its influence on the $\alpha$-effect in mean-field models), we will focus on what the global magnetic helicity can tell us about the global topology of the magnetic field, namely how the toroidal and poloidal fields link with each other and how variations in this linkage manifest themselves, particularly at the outer surface (which can be observed in stars).

Some initial inroads have already been made in this direction in observational studies of global magnetic helicity.  Recent examples of work in this area, which focus on the Sun, include \cite{pipin2014,hawkes2018magnetic,pipin2019evolution}. In particular, \cite{hawkes2018magnetic} show that helicity flux during solar minima is a good predictor of the subsequent solar maxima, and can improve upon predictions based on the polar magnetic field alone.  This result seems to be stronger when the helicity flux measured in only one hemisphere is considered, a calculation that can be performed due to the equatorial symmetry of helicity density at the solar surface. Magnetic helicity has also begun to be measured in stars other than the Sun \citep[e.g.][]{lund2020measuring,lund2021field}. Again, only the surface density of magnetic helicity can be measured, and at limited resolution. Despite this, scaling laws relating the strength of the surface helicity density to that of the toroidal field have been found \citep{lund2021field}. Further, the results of the latter study suggest that the field topology is different for stars on different parts of their evolutionary track. Thus, it may be possible to use magnetic helicity density, in conjunction with other observables, to characterise different stages of stellar, and hence dynamo, evolution.

The purpose of this note is to investigate global magnetic helicity in various types of self-consistent convective dynamos in spherical shells. Our approach for this pilot investigation is not to try to model the Sun or a particular star or planet, but rather to analyse from a general standpoint some typical solutions to a well-studied model for spherical dynamos \citep{Simitev2005,Busse2011} that are representative of various known dynamo regimes. The three particular solutions that we consider increase in complexity and allow us to assess the usefulness of magnetic helicity as a prediction and analysis tool when confronted with increasingly chaotic spatial and temporal flow and field structures. 

The paper is set up as follows. First, the main model equations are introduced, together with brief remarks on the numerical method used. This is followed by a description of magnetic helicity in spherical shells. We then consider each particular dynamo solution in turn, with a focus on what information is provided by magnetic helicity. The report ends with a summary of the results and a discussion of their interpretation.

\section{Main equations}
\subsection{Mathematical model formulation}

Following an established model of convection-driven spherical dynamo action ~\citep[e.g.][]{Busse2006b,Simitev2009a,Simitev2012a} we consider a  spherical shell that rotates about a fixed axis with a constant angular speed $\Omega$. A static state is assumed to exist with the temperature distribution
\begin{subequations}
\begin{gather}
    T_S = T_0-{\beta d^2r^2}/{2} +\Delta T \eta r^{-1}(1-\eta)^{-2},\\
    \beta = q/(3\kappa c_p),\\
   T_0=T_1-\Delta T/(1-\eta), \quad T_S (r_i)=T_1, \quad \quad T_S (r_o)=T_2, \quad \Delta T = T_2-T_1, 
\end{gather}
\end{subequations}
where $T_1$ and $T_2$ are constant temperatures at the inner and outer spherical boundaries, $\eta=r_i/r_o$ is the ratio of the inner radius $r_i$ to the outer radius $r_o$, $q$ is a uniform heat source density, $\kappa$ is the thermal diffusivity, $d$ is the shell thickness and $r$ is the magnitude of the position vector to a point in the spherical shell.

Within the spherical shell, we solve the equations of magnetohydrodynamics (MHD) under the Boussinesq approximation,
\begin{subequations}
\label{convdynamoeqns}
\begin{align}
    \bnab\bdot\uv&=0, \quad \bnab\bdot\Bv=0,\label{solenoidal}\\
    \left(\frac{\partial}{\partial t} + \uv\bdot\bnab\right)\uv &= -\bnab\pi-\tau\kv\times\uv+\vartheta\rv + \nabla^2\uv+ \Bv\bdot\bnab \Bv,\label{momentum} \\
    P\left(\frac{\partial}{\partial t} + \uv\bdot\bnab\right)\vartheta&=[R_i + R_e \eta r^{-3}(1-\eta)^{-2}]\rv\bdot\uv+\nabla^2\vartheta,\label{heat} \\
    P_m\left(\frac{\partial}{\partial t} + \uv\bdot{\bnab} \right)\Bv&=P_m\Bv\bdot\bnab\uv+\nabla^2\Bv.\label{induction}
\end{align}
\end{subequations}
In the above equations, $\uv$ is the velocity field, $\Bv$ is the magnetic field, $\vartheta$ is the deviation from the static temperature distribution, $\pi$ is an effective pressure including all terms that can be represented as a gradient, $\rv$ is the position vector and $\kv$ is the unit vector in the positive vertical direction about which the shell rotates. 

Above, the Boussinesq approximation is used, where the density $\rho$ is constant everywhere except in the buoyancy term. There the density varies linearly with respect to a constant background density $\rho_0$ and has the form
\begin{equation}
\rho=\rho_0(1-\alpha\vartheta),
\end{equation}
where $\alpha$ is a constant and represents the specific thermal expansion coefficient. The non-dimensional parameters in equations \eqref{convdynamoeqns} are
\begin{gather}
    R_i=\frac{\alpha\gamma\beta d^6}{\nu\kappa},\quad R_e=\frac{\alpha\gamma\Delta T d^4}{\nu\kappa},  \quad \tau=\frac{2\Omega d^2}{\nu}, \quad P=\frac{\nu}{\kappa}, \quad P_m=\frac{\nu}{\lambda},
\end{gather}
which are the internal and external thermal Rayleigh numbers $R_i$ and $R_e$, the Coriolis number $\tau$, the Prandtl number $P$ and the magnetic Prandtl number $P_m$, respectively. Of the quantities not yet defined, $\lambda$ is the magnetic diffusivity, $\nu$ is the viscosity and $\gamma$ is a constant related to the gravitational acceleration $\gv= -d\gamma\rv$.

To solve the above equations, the velocity and magnetic field are first split into poloidal and toroidal parts, 
\begin{subequations}
\begin{align}
  \uv &= \uv_P+\uv_T = \bnab\times(\bnab\times \rv v)+\bnab\times  \rv w, \\
    \Bv &= \Bv_P+\Bv_T =\bnab\times(\bnab\times  \rv h)+\bnab\times \rv g,\label{bpt}
\end{align}
\end{subequations}
where $\rv=r\ev_r$. The scalar quantities are decomposed into spherical harmonics,
\begin{subequations}
\begin{align}
    v &= \sum_{l=0}^\infty\sum_{m=-l}^l V_l^m(r,t)P^m_l(\cos\theta){\mathrm e}^{{\mathrm i}m\phi}, \quad w=\sum_{l=0}^\infty\sum_{m=-l}^lW_l^m(r,t)P^m_l(\cos\theta){\mathrm e}^{{\mathrm i}m\phi},\label{exact1}\\ 
    g&=\sum_{l=0}^\infty\sum_{m=-l}^lG_l^m(r,t)P^m_l(\cos\theta){\mathrm e}^{{\mathrm i}m\phi}, \quad h=\sum_{l=0}^\infty\sum_{m=-l}^lH_l^m(r,t)P^m_l(\cos\theta){\mathrm e}^{{\mathrm  i}m\phi}, \label{exact2}
\end{align}
\end{subequations}
where $P^m_l(\bdot)$ denotes the associated Legendre polynomials of first kind and $(r,\theta,\phi)$ are spherical coordinates. Once all the scalars are decomposed as above, the radial functions are expanded in terms of Chebychev polynomials and the series are truncated to a chosen resolution $(n_r,n_\theta,n_\phi)$. The MHD equations are solved using a pseudospectral method and a combination of Crank-Nicolson and Adams-Bashforth time integration schemes.
Further details of the numerical method can be found in \cite{Tilgner1999} and an open source version of the code is available at \cite{Silva2018b}. For brevity, we do not repeat this description and instead guide the reader to these other works. The calculations performed in this work have been run with the resolutions $(n_r,n_\theta,n_\phi)=(33,64,129)$ and $(n_r,n_\theta,n_\phi)=(41,96,193)$. Azimuthally averaged components of the fields $v$, $w$, $h$ and $g$ will be indicated by an overbar. 

To complete the specification of the above mathematical model, we require boundary conditions. We assume fixed temperatures at 
$r=r_i$ and $r=r_o$ 
and consider both cases with no-slip conditions
\begin{subequations}
\begin{equation}
    v=\frac{\partial v}{\partial r}=w=0,
\end{equation}
and cases with stress-free conditions,
\begin{equation}
    v=\frac{\partial^2v}{\partial r^2}=\frac{\partial}{\partial r}\left(\frac{w}{r}\right)=0.
\end{equation}
For the magnetic field, we assume electrically insulating conditions at $r=r_i$ and $r=r_o$. At these locations, the poloidal function $h$ matches the function $h^{(e)}$ which describes the potential fields outside the spherical shell,
\begin{equation}
    g=h-h^{(e)}=\frac{\partial}{\partial r}(h-h^{(e)})=0.
\end{equation}
\end{subequations}

\subsection{Magnetic helicity}

The magnetic field is generated and sustained by thermal convective motions inside the spherical shell ~\citep[e.g.][]{Busse2005a}. The field emanates outside of the spherical fluid shell where, in the absence of sustaining fluid motion, it takes the form of a freely decaying potential field.    
Thus, the magnetic field is not closed and we must consider relative magnetic helicity \citep{bergerfield1984} instead of classical helicity \citep{woltjer1958theorem,moffatt1969degree}. Assuming $h$ is continuous at the inner and outer boundaries, \cite{berger1985structure} showed that the relative helicity in the spherical shell volume $V$ has the appealing form
\begin{equation}\label{helicity}
    H = 2\int_V\Lv h\bdot\Lv g\, \d V = 2\int_V({\rm curl}^{-1}\Bv_P)\bdot\Bv_T\,\d V,
\end{equation}
where $\Lv=-\rv\times\bnab$. Note that the definition of $\Lv$ is chosen to match the particular magnetic field decomposition used in equation (\ref{bpt}). The representation in (\ref{helicity}) can be interpreted in terms of the global mutual linkage of the poloidal and toroidal magnetic fields \citep[see also][]{bergerhornig2018}. This is a generalized form of linkage as the poloidal and toroidal fields occupy the same region of space. However, as will be made clear in subsequent analysis, the linkage interpretation of magnetic helicity will prove very useful. 

By simple vector algebra, we have that $\Bv_T=\Lv g$. There is no simple relation between $\Bv_P$ and $\Lv h$, but, in spherical coordinates, the latter can be expressed as
\begin{equation}
    \Lv h = \left(0,\frac{1}{\sin\theta}\frac{\partial h}{\partial\phi},-\frac{\partial h}{\partial\theta}\right).
\end{equation}
A test to confirm that the equation for the helicity $H$ has been coded correctly is described in the Appendix.

\section{Dynamo solutions}
\looseness=-1
We now investigate the behaviour of global magnetic helicity in various dynamo solutions. Three cases of increasing complexity are considered as summarized in table \ref{table1}.

\begin{table}[t]
\centering
\begin{tabular}{ | c | c | c | c |} 
  \hline
  & Case 1:  & Case 2:   & Case 3:   \\ 
  & Steady dynamo   & Quasi-periodically  & Aperiodically-   \\ 
   &    & reversing dynamo   &  reversing dynamo  \\
  \hline
$\eta$ & 0.35 & 0.4   & 0.4   \\
$R_i$ & 0  & 3.5 $\times 10^6$  & $8.5 \times 10^5$   \\
$R_e$  & $10^5$  &  0 &  0  \\
$\tau$ & 2000  & $3 \times 10^4$  & $3 \times 10^4$ \\
$P$ &  1 & 0.75  & 0.1   \\
$P_m$ &  5 & 0.65  & 1  \\
Boundary  & no-slip,  &  stress-free,  & stress-free,  \\
  conditions   & fixed temperatures, & fixed temperatures,   & fixed temperatures, \\
& insulating outer space & insulating outer space & insulating outer space \\
\hline
\end{tabular}
\vspace{7mm}
\caption{The three dynamo solutions considered in the study.}
\label{table1}
\end{table}

\subsection{Case 1: Steady dynamo}
To begin with, we will consider a laminar dynamo solution that has been used as a benchmark case by the community for validating and testing the accuracy of numerical codes \citep{christensen2001numerical,Matsui2016}. 
The non-dimensional parameters and boundary conditions are listed in the first column of table \ref{table1}. 

\begin{figure}[t]
\centering
   \centering
   \hspace*{-1mm}  
    \subfigure(a){\includegraphics[height=30mm]{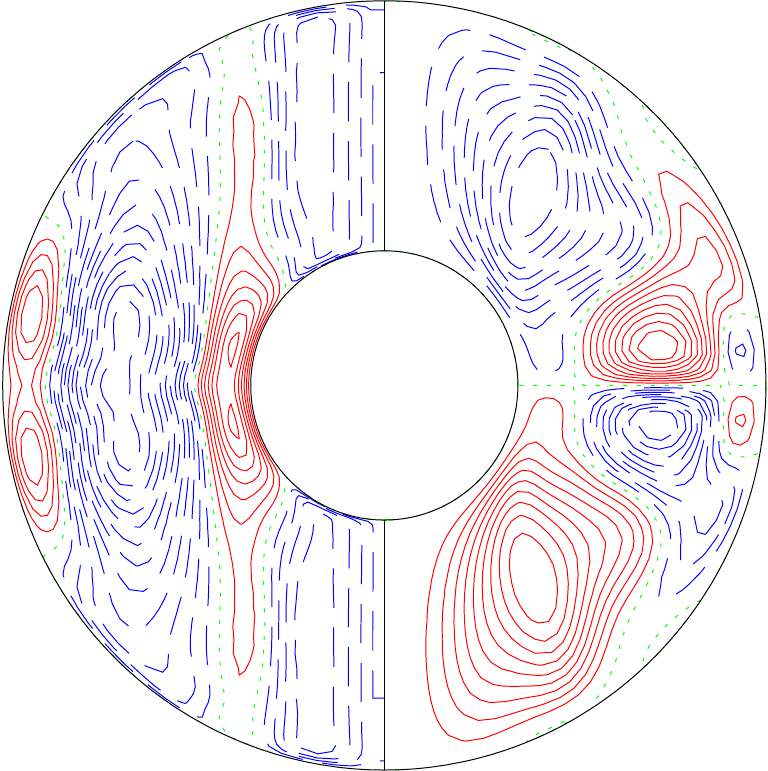}}
    \subfigure(b){\includegraphics[height=31mm]{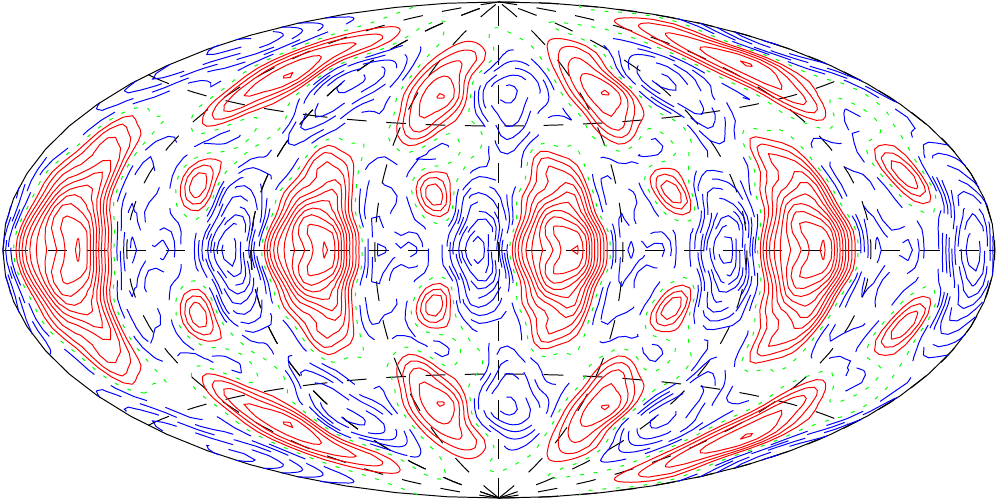}}
    \subfigure(c){\includegraphics[height=30mm]{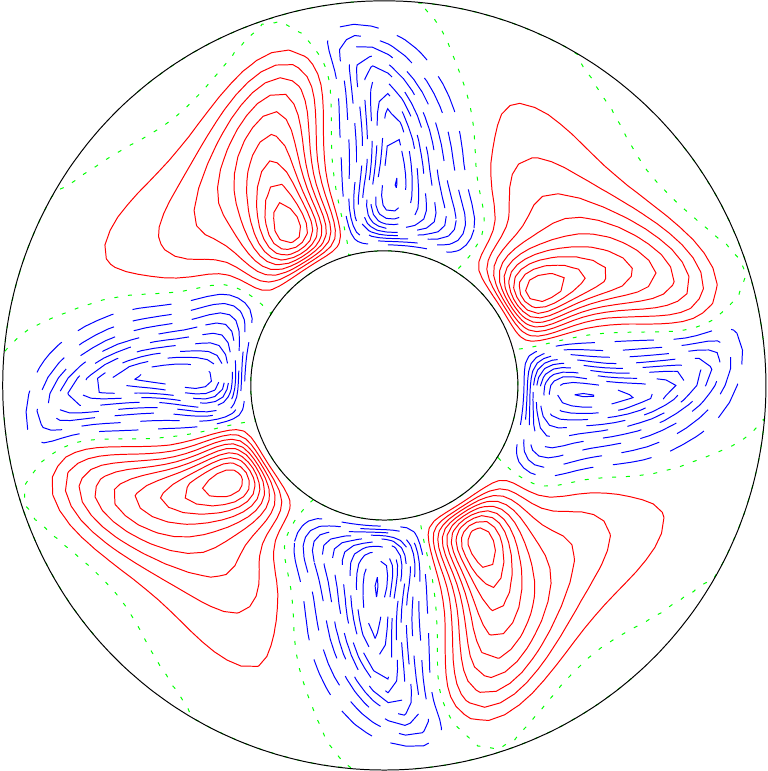}}
\caption{Plots of the velocity field for Case 1 at a particular time (this is a steady flow so other times just represent rigid rotations of this solution) (a) shows lines of constant $\overline{u}_{\phi}$ (left
       half) and $r \sin \theta \dd_\theta \overline{v}$ (right half) in a meridional plane, visualising the toroidal and poloidal fields respectively; (b) shows lines of constant $u_r$ at $r= r_i+0.5$. There is a distinct pattern of columnar convection at both the equator and near the poles; and (c) shows lines of constant $r\partial_\phi v$ in the equatorial plane.  Blue is negative, red is positive and green is zero in this figure and in all subsequent contour plots. (Colour online)}
\label{case1u}
\end{figure}

\begin{figure}[t]
   \centering
   \hspace*{-1mm}  \subfigure(a){\includegraphics[height=29mm]{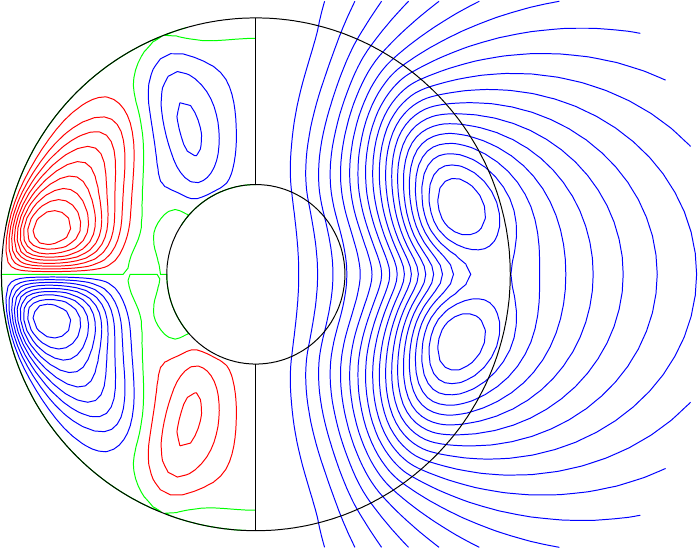}}
    \subfigure(b){\includegraphics[height=30mm]{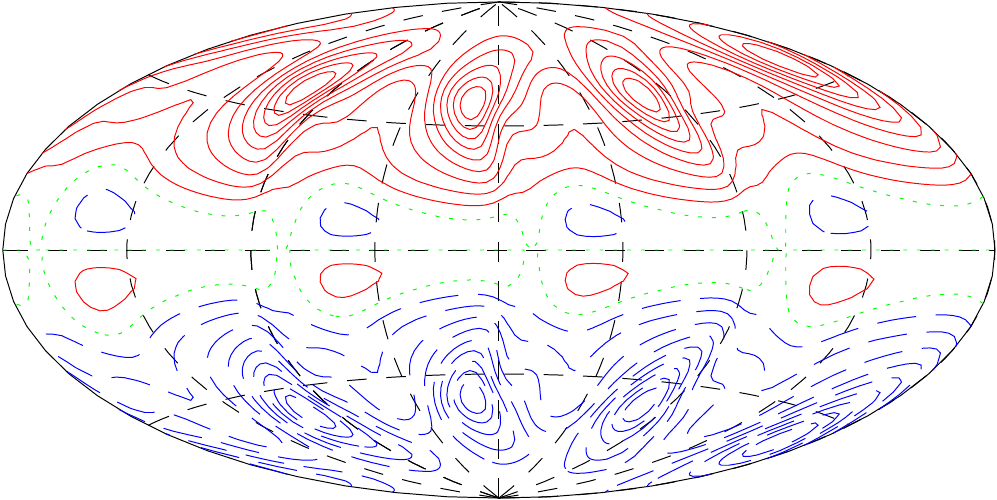}}
    \subfigure(c){\includegraphics[height=29mm]{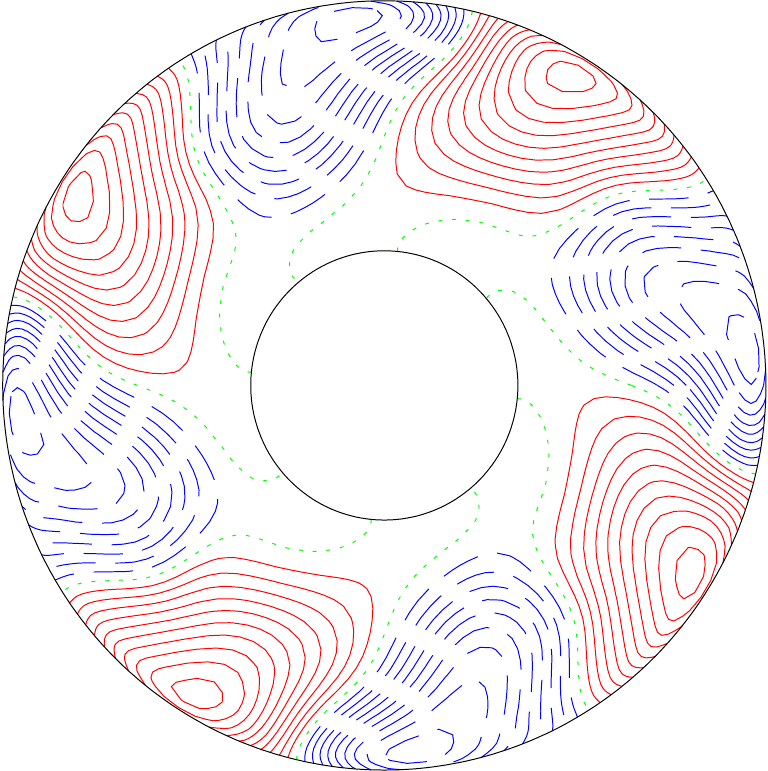}}
    
     \hspace{-2mm}
    \subfigure(d){\includegraphics[height=28mm]{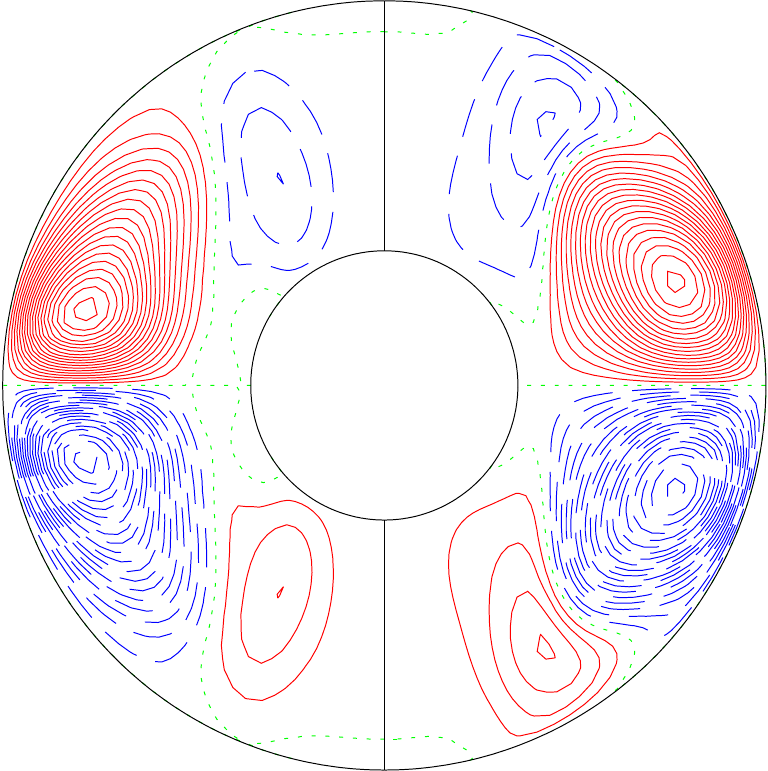}}
    \hspace{8mm}
    \subfigure(e){\includegraphics[height=30mm]{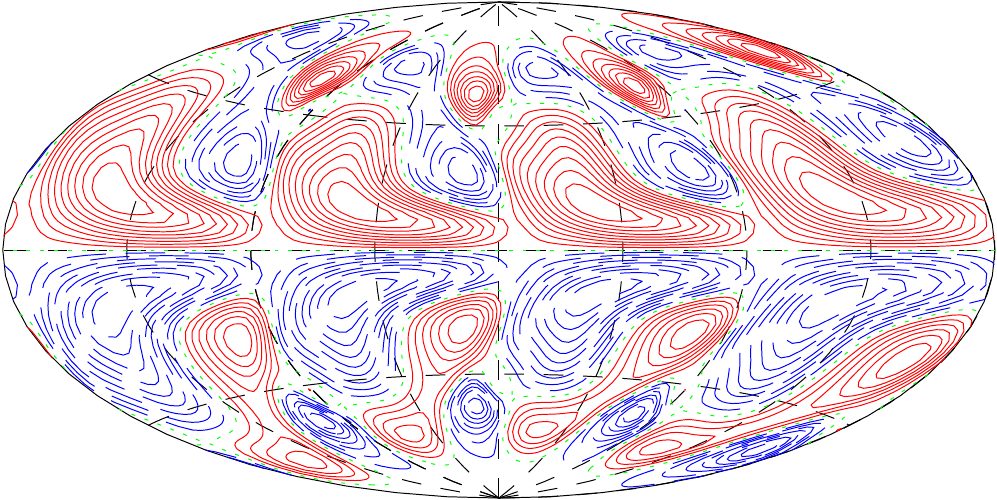}}
    \subfigure(f){\includegraphics[height=29mm]{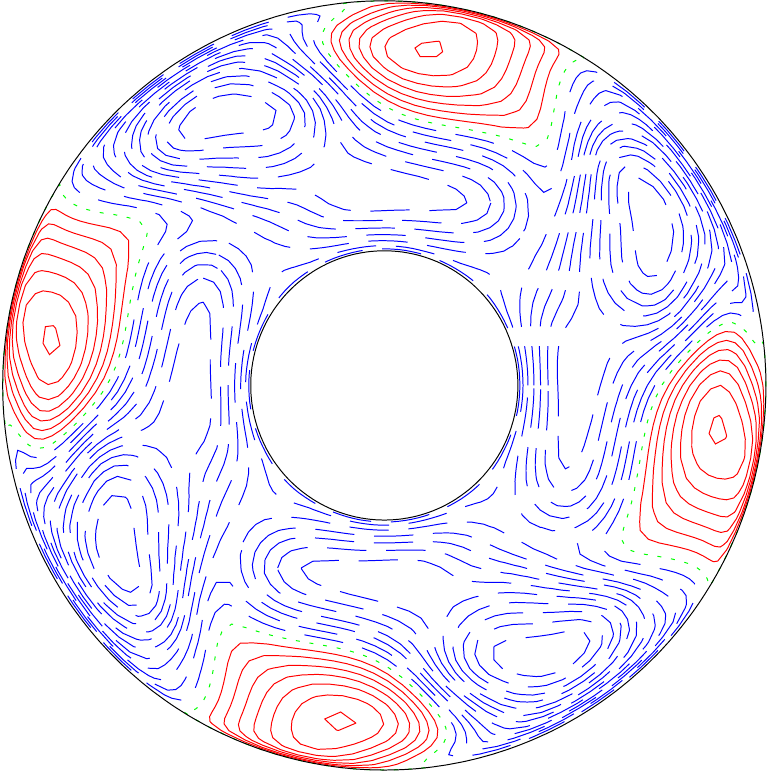}}
     \caption{Plots of the magnetic field (top row) and magnetic
       helicity (bottom row) for Case 1 at a particular time. The top panel (a) shows
       meridional lines of constant $\overline{B}_{\phi}$ (left
       half) and
       $r \sin \theta \dd_\theta \overline{h}$ (right half)
       visualising the toroidal and poloidal fields respectively; (b) shows lines of constant $B_r$ at $r=r_i+0.96$; and (c) displays equatorial streamlines, $r\partial_\phi h$ = const. The bottom panel (d) shows contour lines of azimuthally averaged helicity density (left half) and the helicity density in a particular slice $\phi=\text{const}$ (right half); (e) shows the helicity density at $r=r_i+0.96$; and (f) shows the helicity density at the equatorial plane. (Colour online)}      
    \label{case1plots}
\end{figure}

The main feature of this dynamo is that it develops a steady-state for both the magnetic and velocity fields. The steady velocity profile exhibits columnar convection, as shown in figure \ref{case1u}. Focusing now on the magnetic field, there is a global dipole with inversions at the equator. This situation, together with other details, is displayed in figure \ref{case1plots}.

In figure \ref{case1plots}, the top row reveals the behaviour of the magnetic field in (a) the meridional plane, (b) a near-outer spherical surface, and (c) the equatorial plane. The left-most plot shows the toroidal and poloidal components of the magnetic field. Both quantities shown in the hemispheres are meridional averages. In the left-hand hemisphere, displaying the large-scale behaviour of the toroidal field, the magnetic field pattern is perfectly antisymmetric about the equator. In the right-hand hemisphere, revealing the large-scale behaviour of the poloidal field, a dominant dipolar field is present, except near the surface at the equator. Here, smaller and weaker (compared to the polar fields) positive and negative polarities exist, as can be seen from the surface radial magnetic field plot in (b). The radial component of the magnetic field shows a dipolar symmetry, which means $B_r = 0$ on the equatorial plane. This magnetic profile, which rotates rigidly with the sphere, is simple enough to provide a complete interpretation of the behaviour of magnetic helicity. 

\begin{figure}[t]
    \centering
    \subfigure(a){{\includegraphics[height=35mm]{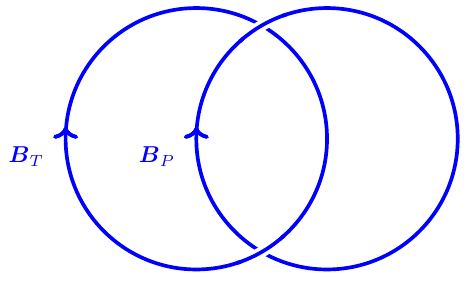}}} \subfigure(b){{\includegraphics[height=35mm]{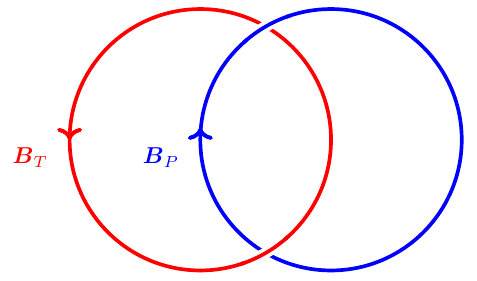}}}
    \caption{A representation of the linkage of toroidal and poloidal fields. Colours correspond to those in figure \ref{case1plots}(a). The  (Gauss) linkage in (a) is $-1$ and in (b) is $+1$. These signs correspond with those of the helicity density in figure \ref{case1plots}(d). (Colour online)}
    \label{linkage}
\end{figure}

The bottom row of figure \ref{case1plots} displays the contour lines of (d) azimuthally averaged helicity density (left half) and unaveraged helicity density in a particular slice (right half), (e) the helicity density at the outer surface, and (f) the helicity density in the equatorial plane. 
Figure \ref{case1plots}(d) reveals how the sign of the helicity density in different parts of the spherical shell depends on the linkage of the toroidal and poloidal fields. In order to clarify the signs of the helicity density displayed in this figure, we sketch an idealized linkage of the toroidal and poloidal fields in figure \ref{linkage}.
The representation of linkage in figure \ref{linkage} is idealized because the toroidal and poloidal fields are space-filling in the spherical shell and are not the union of two disjoint regions. However, due to the relative simplicity of the field structure (clearly separated regions of positive and negative toroidal field), this idealization serves to indicate the correct sign of linkage. Figure \ref{linkage}(a) shows the linkage of a negative (blue) part of the toroidal field with the poloidal field and figure \ref{linkage}(b) shows the linkage of a positive (red) part of the toroidal field with the poloidal field. The Gauss linking numbers for these links are $-1$ and $+1$ respectively, corresponding to the negative and positive patches of helicity density in figure \ref{case1plots}(d). Just to reiterate, this is only a simple representation of the linkage, but it is effective in showing how the magnetic helicity reveals information about the connectivity of poloidal and toroidal fields on large scales.

Figure \ref{case1plots}(d) displays the distribution of helicity density near the surface. This shows a more detailed morphology compared to the azimuthal average plots, but there is still a clear dominance of positive helicity density in the north and negative helicity density in the south. As in the other plots, there is perfect antisymmetry about the equator. Thanks to this property, together with this being a steady solution, $H=0$.

This simple dynamo case is a clear example showing that although the total magnetic helicity is zero, this value is due to the cancellation of positive and negative large-scale structures. In particular, the signs of these structures depend on the linkage of poloidal and toroidal fields. Magnetic helicity, however, is not only a measure of linkage but is weighted by magnetic flux. This consideration will become important for understanding cases exhibiting more irregular temporal behaviour and spatial morphology as discussed below.

\subsection{Case 2: Quasi-periodically reversing dynamo}
We now consider a time-dependent quasi-periodic dynamo solution that regularly reverses the signs of both the poloidal and toroidal parts of the magnetic field. The non-dimensional parameters for this case, together with the boundary conditions, are listed in the second column of table \ref{table1}. 

\begin{figure}[t]
    \centering
    \subfigure(a){\includegraphics[height=30mm]{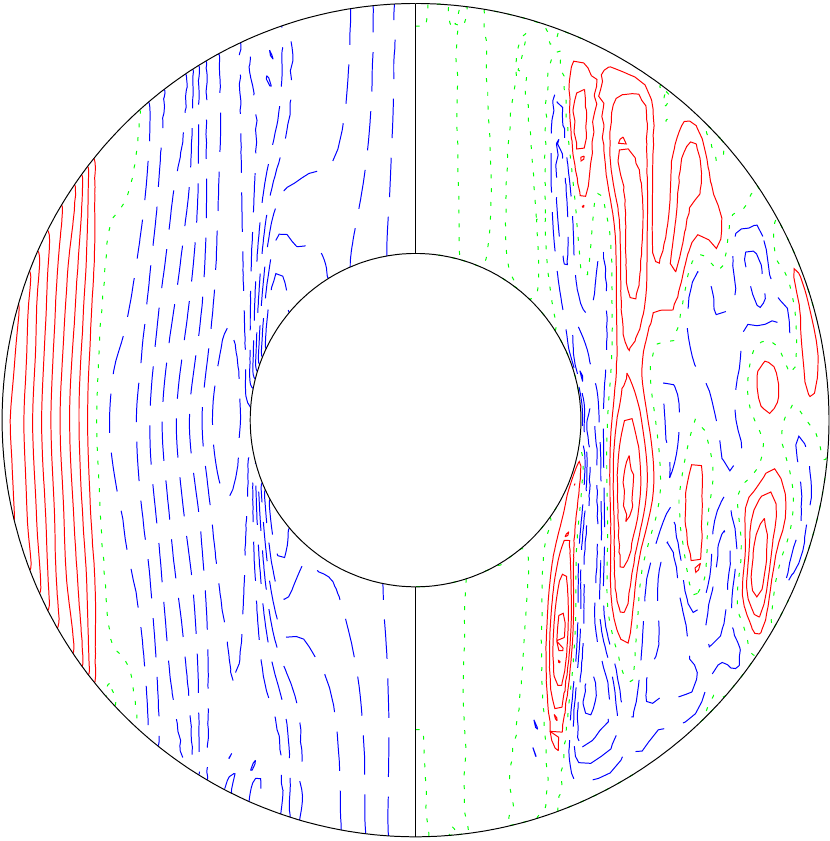}}
    \subfigure(b){\includegraphics[height=32mm]{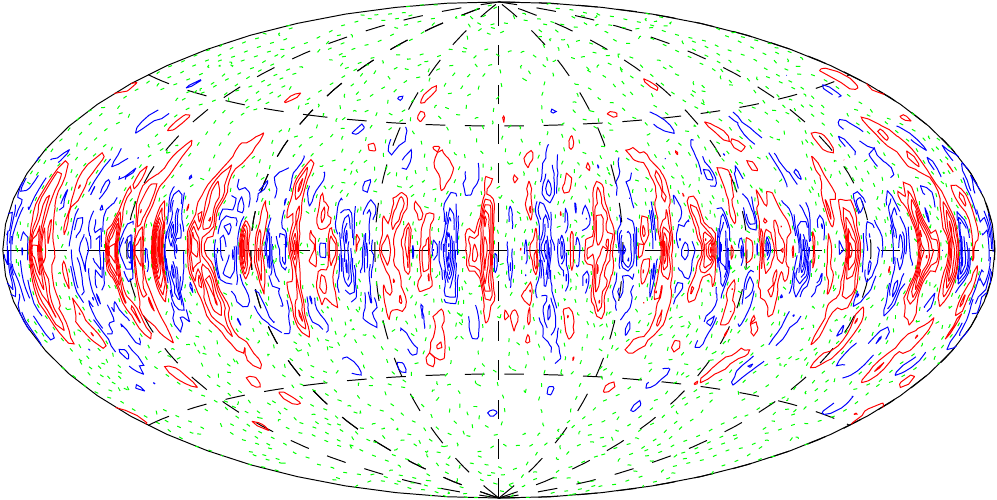}}
     \subfigure(c){\includegraphics[height=30mm]{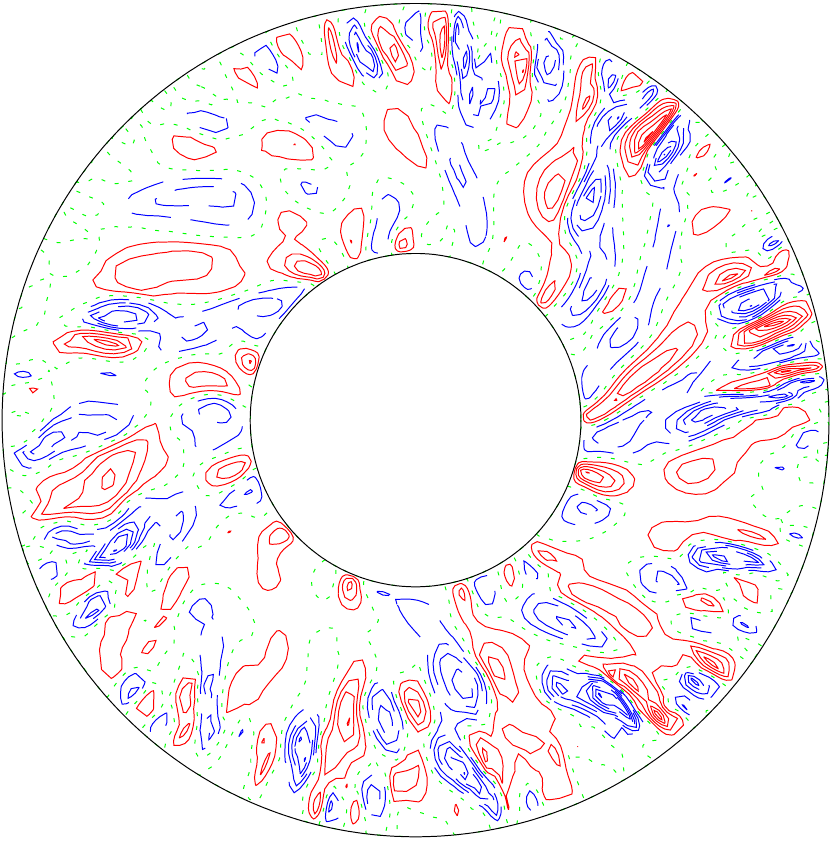}}
    \caption{Typical structures of the velocity field for Case 2. (a) shows lines of constant ${\overline{u}_\phi}$ in the left half and streamlines $r \sin \theta \partial_\theta \overline{v}$ = const in the right half, all in the meridional plane; (b) shows lines of constant $u_r$ at $r= r_i+0.5$; and (c) shows streamlines, $r\partial_\phi v$ = const. in the equatorial plane. These images correspond to the time $t = 5.538$ in the simulation. (Colour online)}
    \label{case2slices}
\end{figure}

Figure \ref{case2slices} displays the typical structure of the velocity field. Although there is columnar convection like Case 1, the velocity is now evolving chaotically in time and exhibits much smaller-scale structures.

\begin{figure}[t]
    \centering
    \subfigure(a){\includegraphics[scale=0.4]{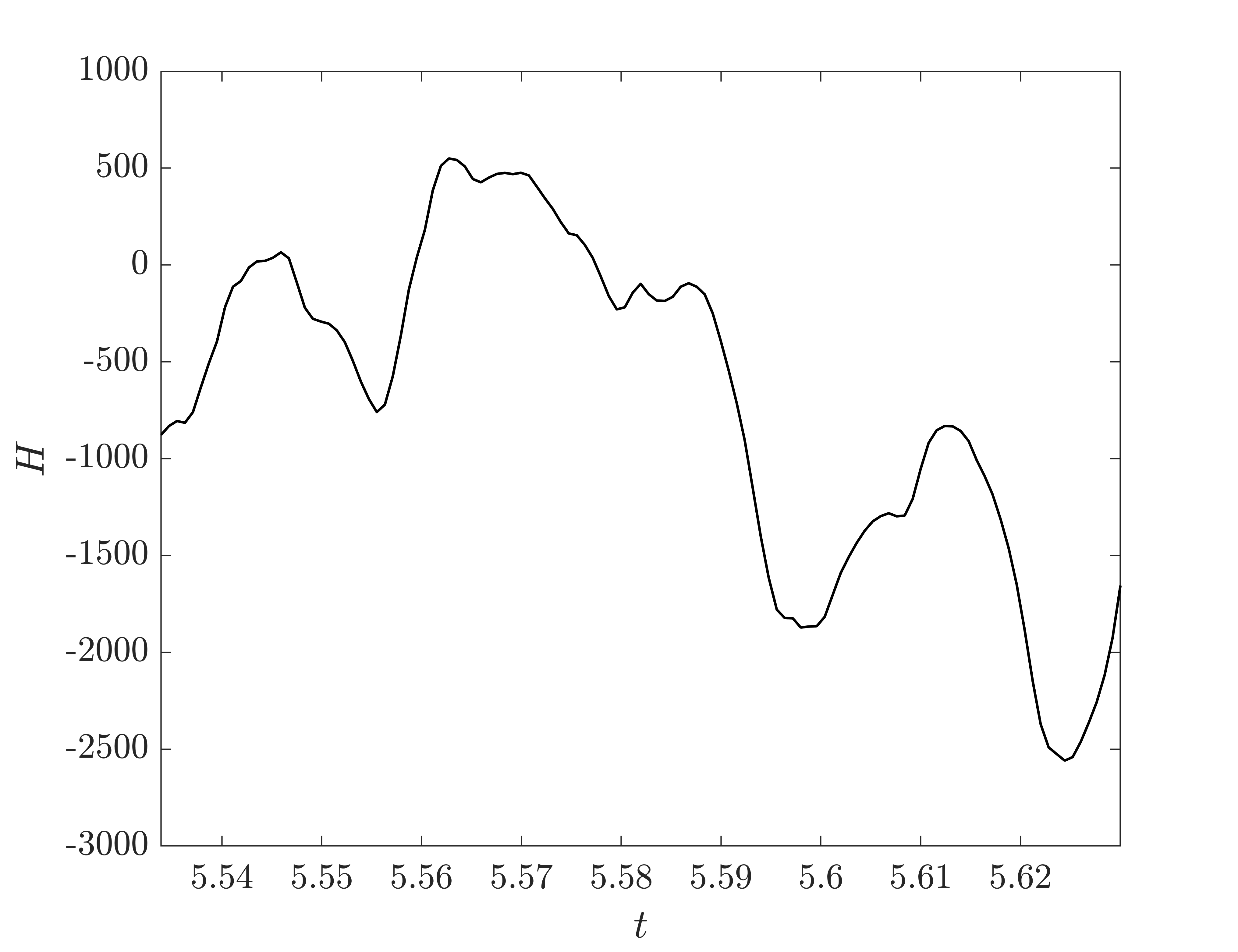}}\subfigure(b){\includegraphics[scale=0.4]{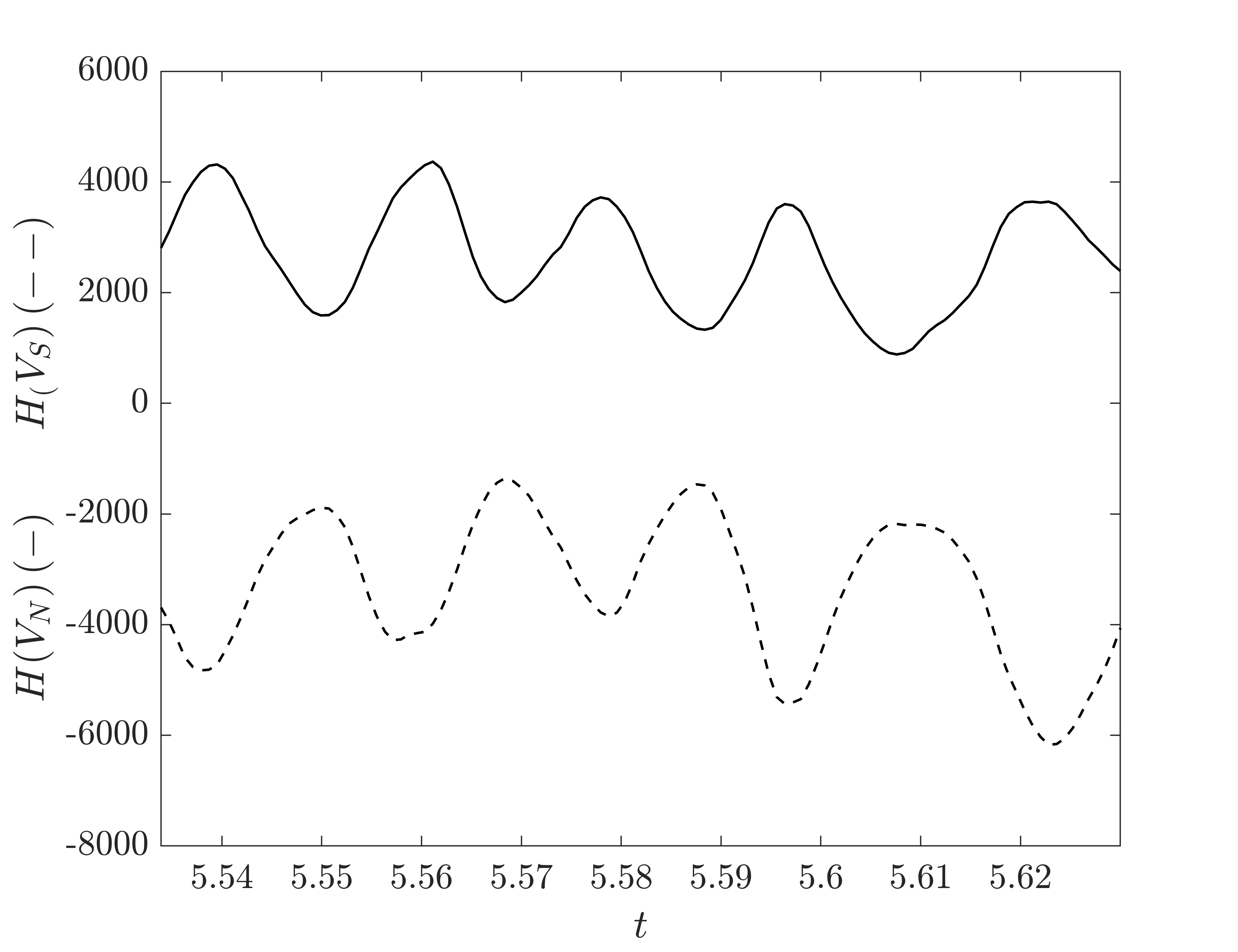}}
    \caption{Time series of magnetic helicity for Case 2. (a) shows the total helicity as a function of time. (b) shows the helicity density integrated in the northern hemisphere (solid) and the southern hemisphere (dashed).}
    \label{hel_time_ser}
\end{figure}

Unlike in Case 1, where the field structure was fixed in time and interpreting the helicity density in terms of linkage was straightforward, here the situation is more involved. Further, the magnetic Prandtl number $P_m$ is an order of magnitude smaller for this case compared to Case 1, so the total magnetic helicity is not expected to be strongly conserved. This is indeed the case upon examination of figure \ref{hel_time_ser}(a).

\begin{figure}[t]
    \centering
    \subfigure(a){\includegraphics[scale=0.4]{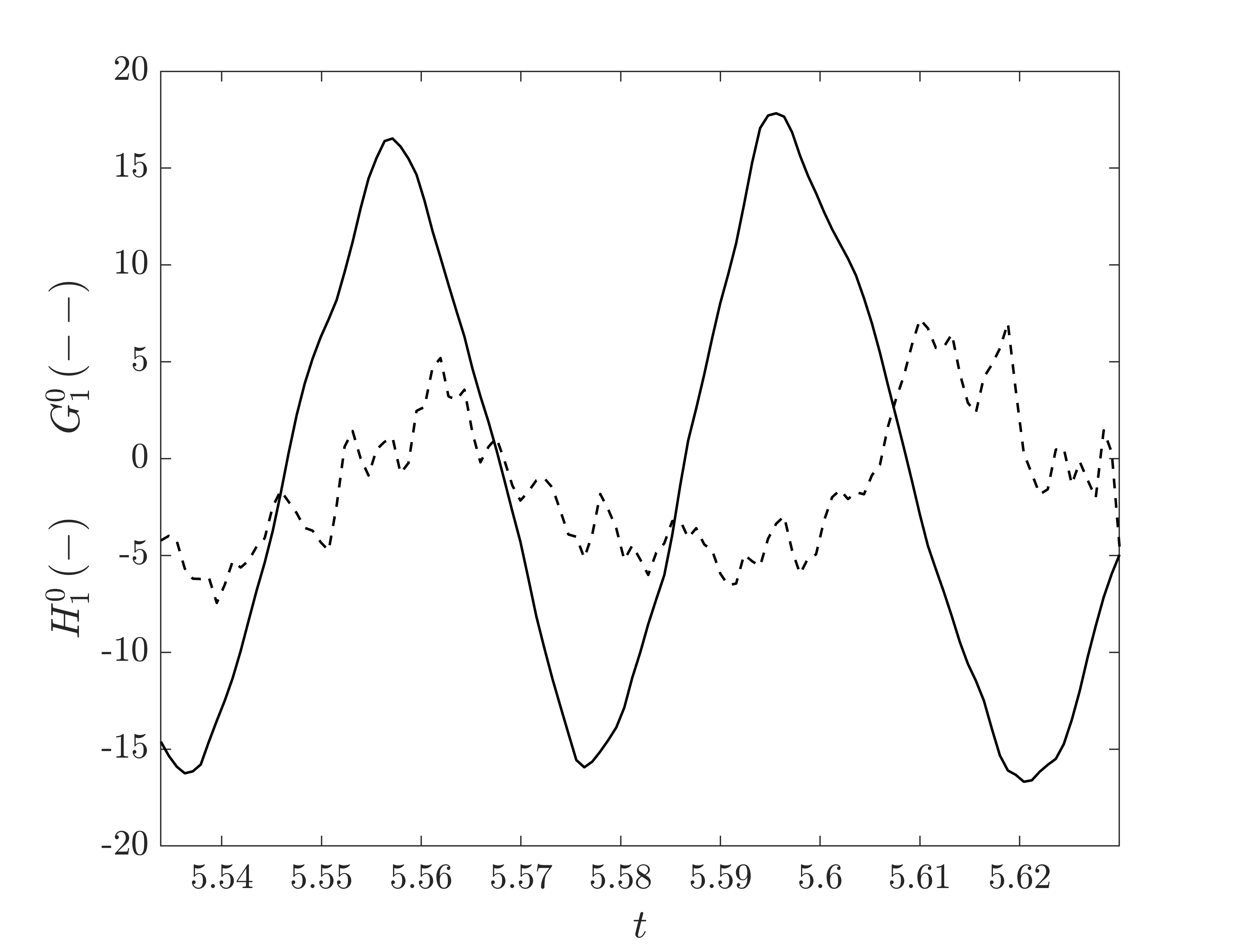}}\subfigure(b){\includegraphics[scale=0.4]{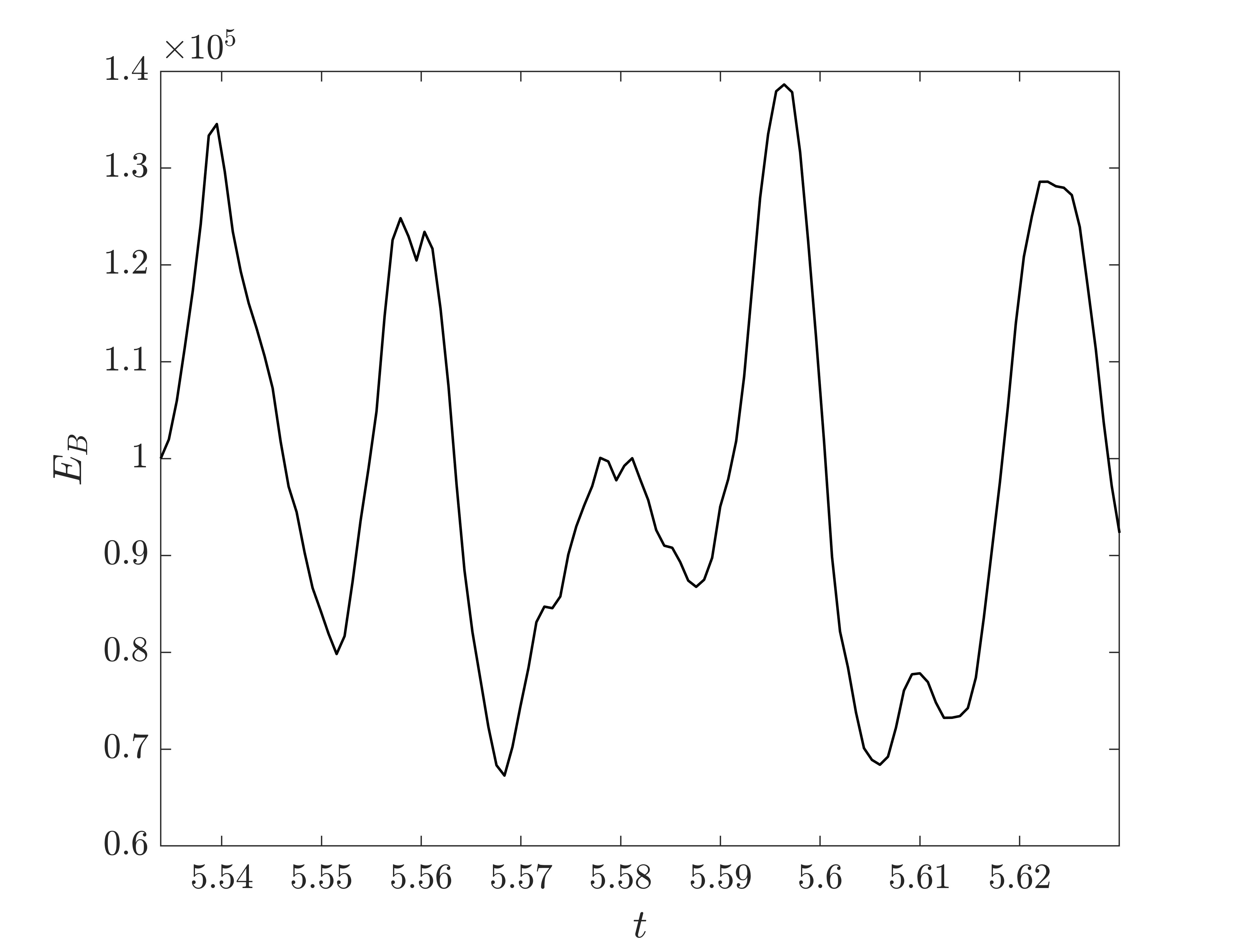}}
    \caption{Time series related to reversals. (a) displays reversals of the poloidal and toroidal fields. The poloidal and toroidal fields are represented by their dominant spectral expansion components $H^0_1$ (solid line) and $G^0_1$ (dashed line), respectively. Both of these quantities are calculated at $r=r_i+0.5$. (b) displays the total magnetic energy $E_B$, which oscillates at a frequency similar to the hemispheric helicities displayed in figure \ref{hel_time_ser}.}
    \label{pt_per_rev}
\end{figure}

Since magnetic helicity is not conserved in time for this dynamo, we cannot make any clear causal conclusions based on helicity conservation. However, this does not mean that magnetic helicity does not reveal useful information. To reveal the structure of magnetic helicity in this dynamo solution, we restrict its calculation to northern and southern hemispheres separately. The results are shown in figure \ref{hel_time_ser}(b). There is a clear wave solution in both hemispheres, with a complete cycle taking about 0.02 time units. This result indicates that although there can be local changes in the magnetic helicity density in a hemisphere, there is no complete change in the linkage of hemispheric toroidal and poloidal fields. This global preservation of linkage arises from the nearly co-temporal reversal of the global toroidal and poloidal fields, leaving little time for a different large-scale linkage to develop. The reversals of the poloidal and toroidal fields are indicated in figure \ref{pt_per_rev}(a). Since the global toroidal and poloidal fields reverse together, the linkage, and thus the overall sign, of magnetic helicity in each hemisphere remains the same.

\begin{figure}[t]
    \centering
    \includegraphics[scale=0.5]{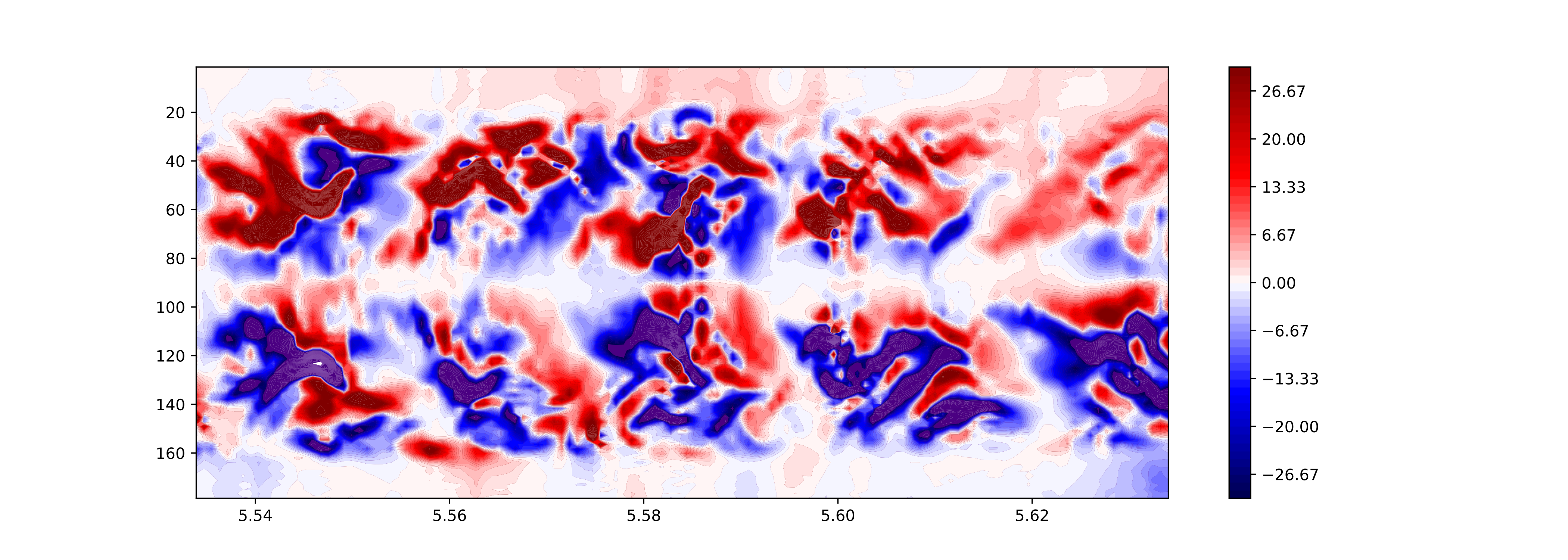}
    \caption{A latitude vs time plot of the magnetic helicity density at the surface for Case 2 with a nonlinear colourmap:
indigo (-400,-30); maroon (-30, 30) and seismic (blue and red) (30,400). (Colour online)}
    \label{lat_per}
\end{figure}

Although the integrated magnetic helicity density does not change sign in each hemisphere, it does oscillate. Magnetic helicity can change due to variations in field linkage and magnetic field strength, the latter indicated by the magnetic energy shown in figure \ref{pt_per_rev}(b), and both these effects cause the oscillation shown in figure \ref{hel_time_ser}(b). The global magnetic helicity, in each hemisphere or in the entire spherical shell, is not an observable quantity. The magnetic helicity density at the surface is observable, however, and, for this dynamo solution, provides an indication of when a reversal develops. Figure \ref{lat_per} displays a time-latitude plot of the magnetic helicity density near the surface.

At $t=5.54$, there is the clear positive/negative split in the north/south hemispheres. This time corresponds to the peaks of the magnetic helicity magnitudes in the hemispheres (see figure \ref{hel_time_ser}(b)). As the hemisphere magnitudes decrease to their minima, this is represented in figure \ref{lat_per} by, first, a more mixed pattern of linkage (i.e. increased negative magnetic helicity density in the north and \emph{vice versa} in the south), and then by a decrease in the strength of the magnetic helicity density. These two behaviours can be seen just before and after $t=5.55$ in figure \ref{lat_per}, and repeat for all the cycles shown. The weakening of the density corresponds to the time when the poloidal and toroidal fields reverse and this is followed by the return of these fields to their peak values, resulting in a new phase of dominating positive/negative magnetic helicity density in the north/south hemispheres.

\section{Case 3: Aperiodically-reversing dynamo}
\begin{figure}[t]
    \centering
    \subfigure(a){\includegraphics[height=30mm]{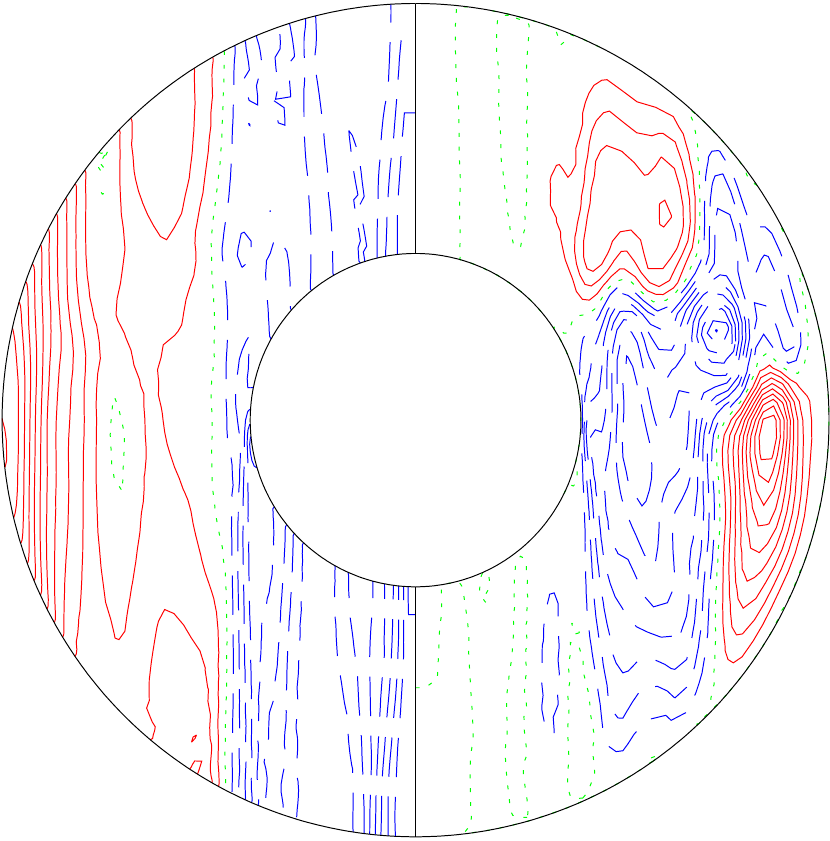}}
    \subfigure(b){\includegraphics[height=32mm]{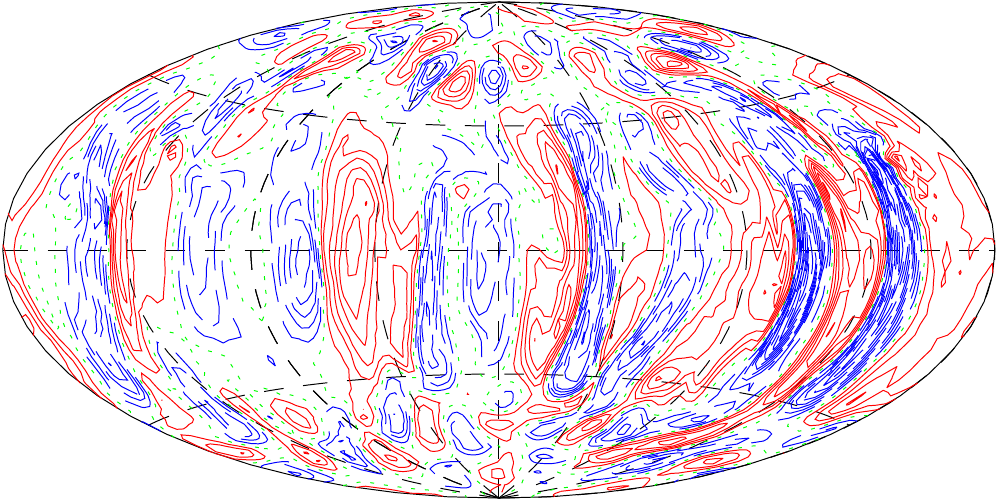}}
     \subfigure(c){\includegraphics[height=30mm]{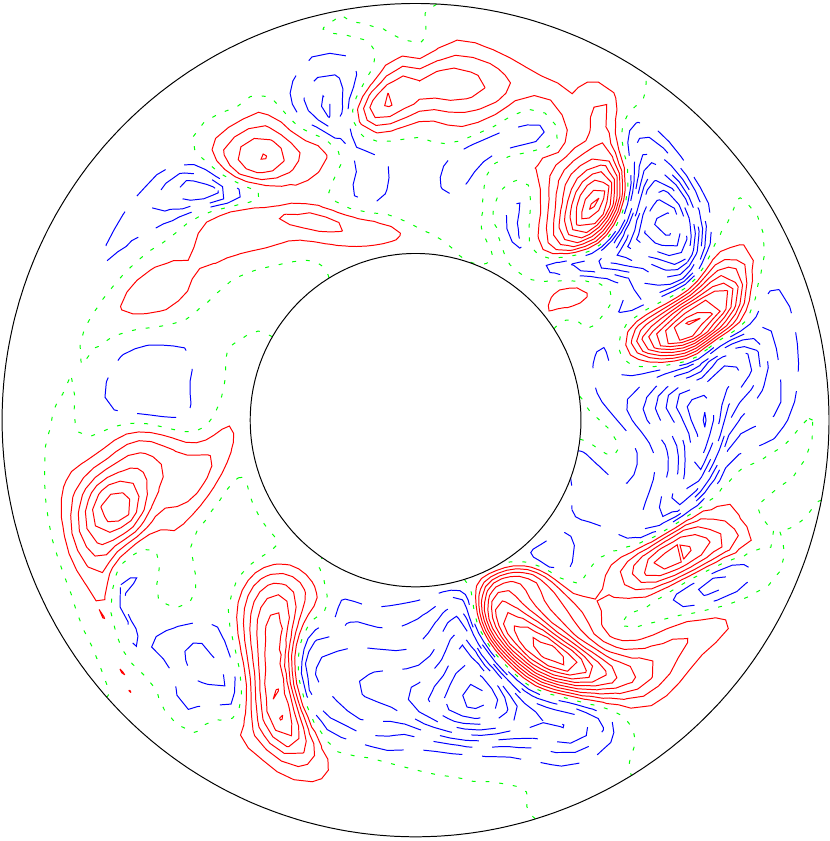}}
    \caption{Typical structures of the velocity field in the case. (a) shows lines of constant ${\overline{u}_\phi}$ in the left half and streamlines $r \sin \theta \partial_\theta \overline{v}$ = const. in the right half, all in the meridional plane; (b) shows lines of constant $u_r$ at $r= r_i+0.5$ and (c) shows streamlines, $r\partial_\phi v$ = const. in the equatorial plane. These images correspond to the time $t = 13.52$ in the simulation. (Colour online)}
    \label{aperiod_vel}
\end{figure}

We now consider a dynamo solution in which aperiodic reversals of the global magnetic field occur. The general behaviour of this dynamo is rather more chaotic compared to the previous two cases considered, and there is no clear precursor pattern for global reversals. The non-dimensional parameters and boundary conditions used for this case are listed in the third column of table \ref{table1}.
For completeness, we show, in figure \ref{aperiod_vel}, typical profiles of velocity components for this dynamo solution. Like Case 2, the velocity has a columnar but chaotic morphology.

\begin{figure}[ht!]
    \centering
    \subfigure(a){\includegraphics[scale=0.40]{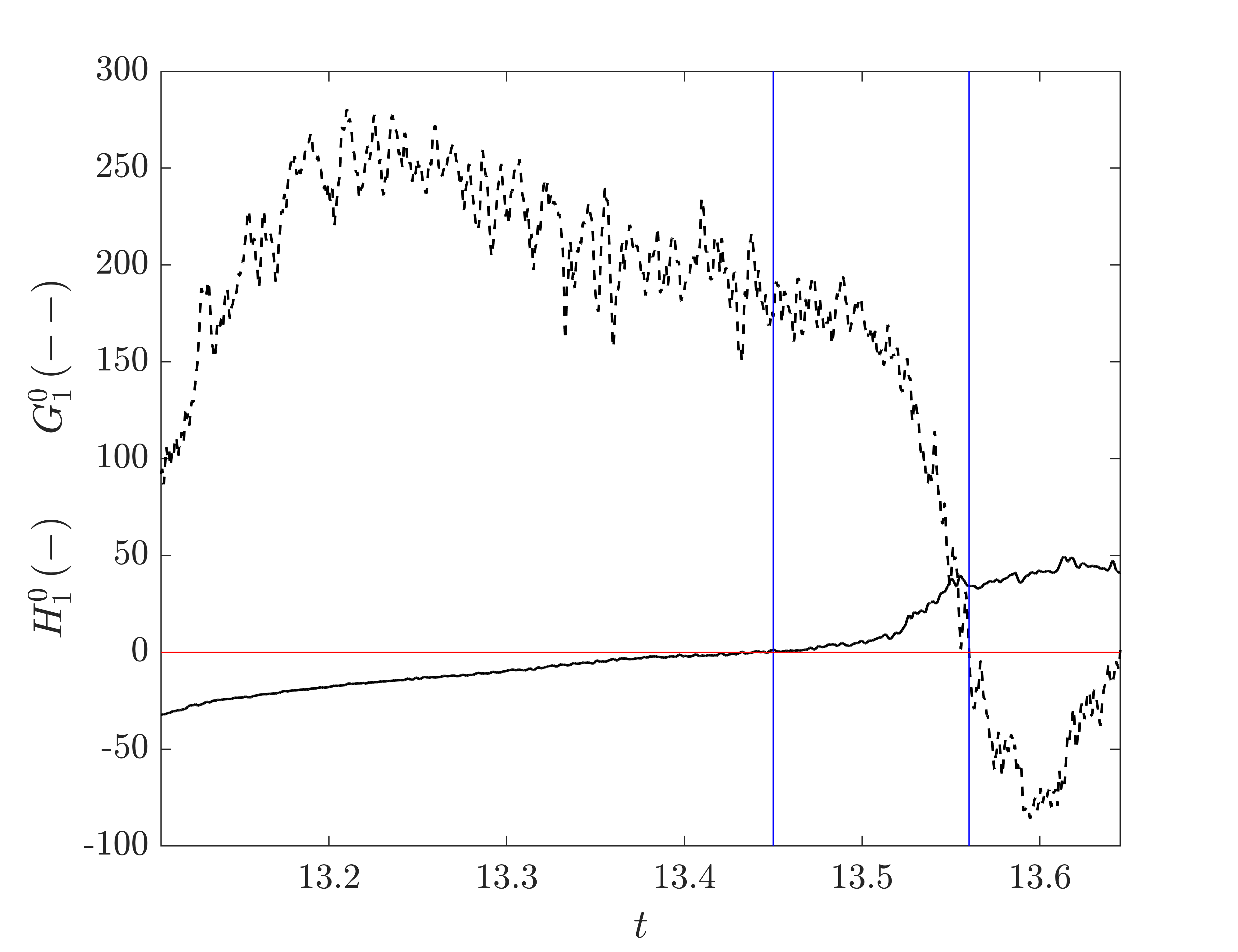}}\subfigure(b){\includegraphics[scale=0.40]{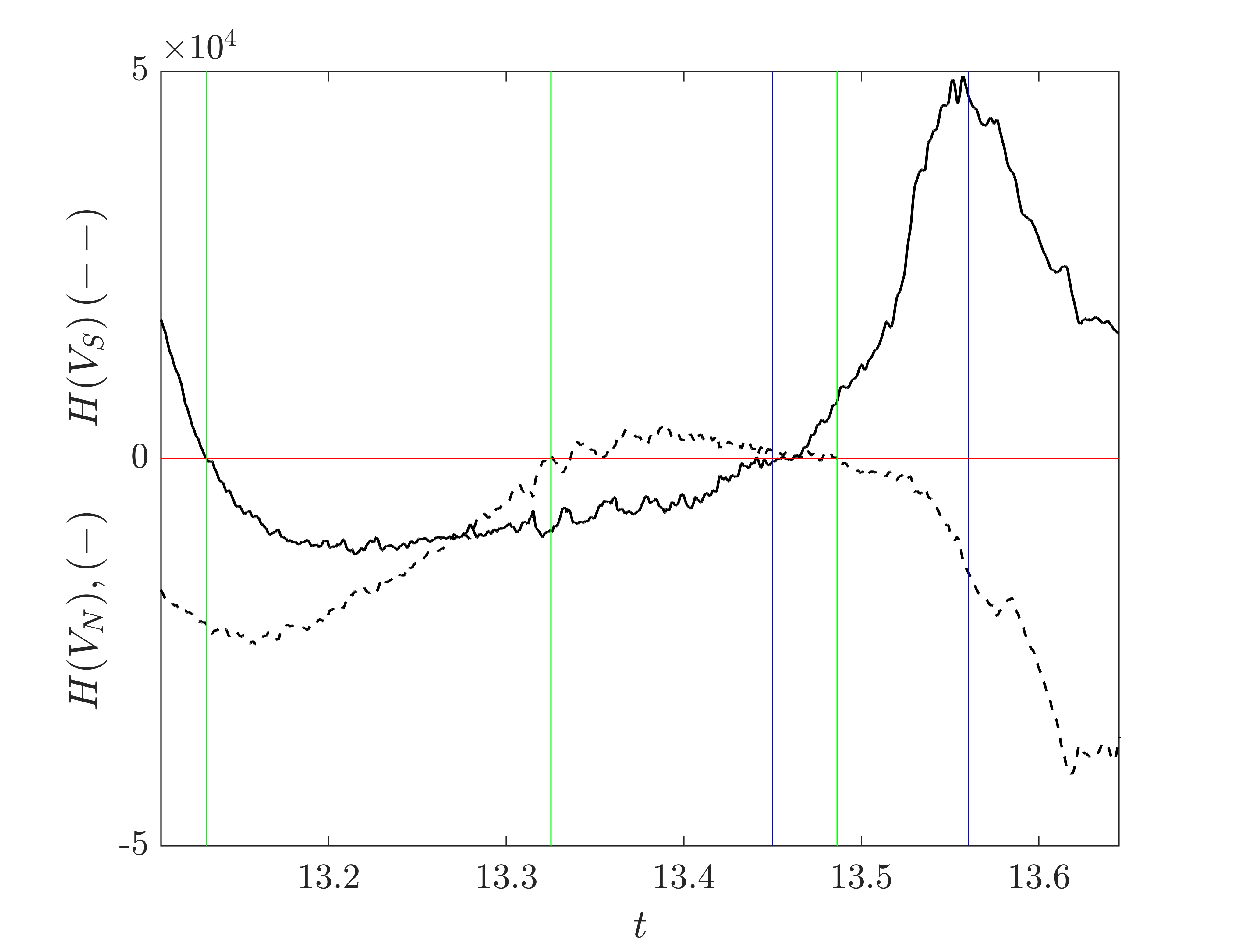}}
    \caption{Time series during a reversal. $H^0_1$ (solid) and $G^0_1$ (dashed) are displayed in (a). Both quantities are evaluated at $r=r_i+0.5$ The blue vertical lines indicate the reversal times of these quantities. The northern (solid) and southern (dashed) helicities are shown in (b). Reversal times are indicated by green vertical lines. Just before the final green line, there are very small reversals about the $H=0$ axis, and these are not shown. (Colour online)}
    \label{chaotic_ts}
\end{figure}

This particular dynamo solution has been studied in \cite{BUSSE2008}. Therefore, rather than repeating the description given in that work, we now focus on the behaviour of magnetic helicity at a reversal. Unlike Case 2, the reversals for this dynamo solution do not occur at regular intervals. That being said, despite the differences between these two cases, the magnetic helicity can again be used to interpret the behaviour of the global magnetic field during a reversal. Figure \ref{chaotic_ts}(a) displays dominant components of the poloidal and toroidal scalars $H^0_1$ and $G^0_1$ at a reversal and figure \ref{chaotic_ts}(b) displays the integrals of the magnetic helicity density in both the north and south hemispheres.  

Apart from the lack of periodicity, the other striking difference compared to Case 2 is that the hemispheric helicities change sign. The blue lines in Figures \ref{chaotic_ts}(a) and (b) indicate the reversals of $H^0_1$ and $G^0_1$ (both evaluated at $r=r_i+0.5$). The first two green lines in figure \ref{chaotic_ts}(b) indicate the reversals of $H(V_N)$ and $H(V_S)$. The helicity reversals occur long before those of $H^0_1$ and $G^0_1$. The third green line in figure \ref{chaotic_ts}(b) marks the time at which the signs of both the hemispheric helicities return to their original values (from the start of the time series). This occurs after the reversal of $H^0_1$ and before that of $G^0_1$. Some care is required in interpreting the reversal times as the helicities are dependent on the magnetic field throughout an entire hemisphere, whereas the values $H^0_1$ and $G^0_1$ presented here are evaluated at a single radius. Despite this, however, the results indicate that an oscillation in the hemispheric helicities, resulting in sign changes, precedes the global reversal. In other words, the linkage of the poloidal and toroidal fields changes, flipping and returning to its original state, leading to a global reversal. This suggests that the topology of the magnetic field changes and reaches an unstable state, in the sense of this state not lasting a long time. The way the magnetic field returns to a stable (longer lasting) configuration, and thus a stable topological state, is through a reversal. This pattern of helicity reversing before a global reversal is also true for the other reversals (labelled 2 and 3) that occur in the solution (for the time span we have simulated it). These results are displayed in table \ref{table}.

\begin{table}[ht!]
\centering
\begin{tabular}{ | c | c | c | c | c | c | c | } 
  \hline
  Reversals & $H^0_1$ & $G^0_1$ & $H(V_N)$ & $H(V_S)$ & $H(V_N)$ and $H(V_S)$  \\ 
  \hline
2 & 26.3232 & 26.374   & 26.199  & 26.3313  &  26.3504 \\
3 & 31.2784  & 31.3534  & 31.169  & 31.2532 & 31.3047 \\
  \hline
\end{tabular}
\vspace{7mm}
\caption{The helicity and magnetic field reversal times for the other observed reversals.}
\label{table}
\end{table}

\begin{figure}[ht!]
    \centering
    \includegraphics[scale=0.4]{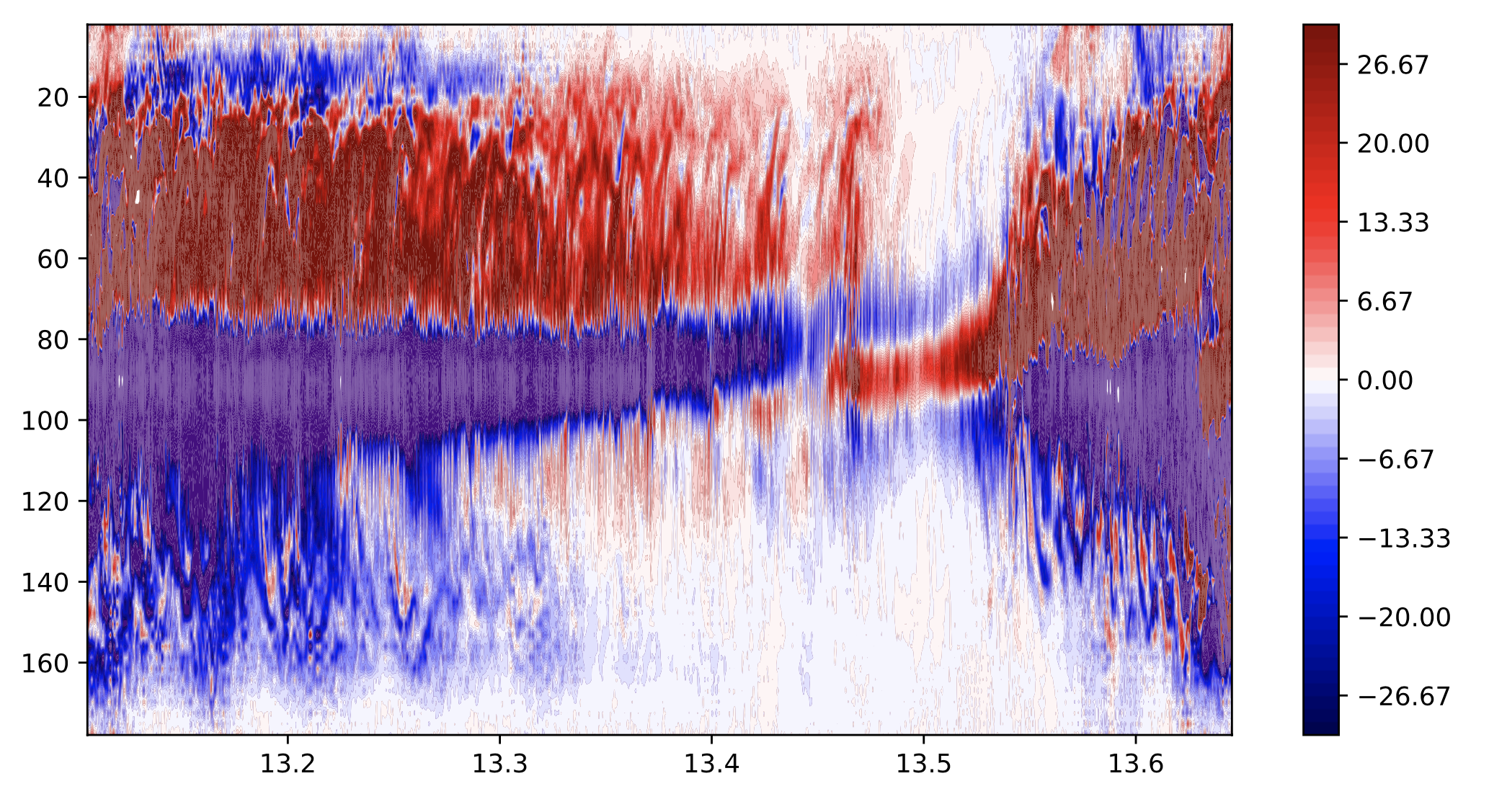}
    \caption{A latitude vs time plot of the magnetic helicity density at the surface for Case 3 with a  nonlinear colourmap: indigo (-500,-30); maroon (-30, 30; shown) and seismic (blue and red) (30,500). (Colour online)}
    \label{lat_chaotic}
\end{figure}

The quantities discussed so far in this section are available in simulations but not to observers, who only have access to data at the outer surface. Figure \ref{lat_chaotic} displays the magnetic helicity density near the outer surface in a latitude-time plot. 

In figure \ref{lat_chaotic} there is also an indication in the behaviour of the magnetic helicity density, at the surface, that reversals are taking place. The reversal of $H^0_1$ in figure \ref{chaotic_ts}, indicating the reversal of the poloidal field, occurs at $t\approx13.45$. From about $t\approx13.28$, there is a signature of stronger patches of positive magnetic helicity density near the equator. At around $t\approx13.45$, there is an abrupt change with much weaker magnetic helicity density at all latitudes. This then switches to a phase of strong positive/negative magnetic helicity density, just southƒnorth of the equator. This signature (a change in sign) is indicative of the reversal of the poloidal field but not the toroidal. When the toroidal field reverses at $t\approx13.552$, the magnetic helicity density distribution returns to its original state. Therefore, even with just information about the magnetic helicity density at the surface, it is possible to identify the reversals of both the global poloidal and toroidal fields.

\section{Discussion}
In this note, we have analyzed three dynamo solutions in rotating spherical shells and have investigated the behaviour of magnetic helicity in each. Our approach has been to adopt a well-studied model (Boussinesq MHD) for convection-driven dynamos in rotating spherical shells and consider dynamo solutions of increasing complexity -- steady (Case 1), periodically-reversing (Case 2), and aperiodically-reversing (Case 3). Although magnetic helicity is not conserved in the final two cases, it does provide important information about the magnetic field in all cases. It has been shown, as indicated by its definition, that magnetic helicity can provide clear information about the linkage of toroidal and poloidal magnetic fields. For the cases with reversals, we have described how magnetic helicity relates to the reversal of both poloidal and toroidal magnetic fields and that changing states of the magnetic helicity density on the surface (an observable quantity in stars) can indicate the onset of reversals. 

Since magnetic helicity is not strictly conserved in the simulations we have considered, we cannot claim that reversals are linked causally to magnetic helicity. However, what our results do suggest is that there are perhaps preferred states of global magnetic linkage for particular dynamo solutions. When the dynamo moves away from such states, it can return via a global reversal. For example, if there is a reversal of the large-scale toroidal field, this changes the global field linkage. To return to the original linkage, the poloidal field reverses, thus completing the global magnetic field reversal and returning the magnetic helicity distribution to what it was before any reversal took place. Therefore, even if magnetic helicity is not the cause of reversals in these simulations, it is a clear signature of the global poloidal-toroidal linkage, which is intimately linked to reversals. Furthermore, maps of the magnetic helicity density at the surface, which are measurable quantities in stellar observations, indicate the onset of reversals. This result is robust for dynamo solutions with different reversal mechanisms. 

As mentioned previously, the main focus on magnetic helicity in dynamos has been the role it plays in the $\alpha$-effect, in mean-field models. Thus, there is an assumed scale separation of magnetic helicity. Our simulations focus predominantly on the large-scale magnetic helicity, which alone (without the small-scale contribution) is not conserved. Nevertheless, our results do tie in qualitatively with some results from mean-field studies. For example, in a dynamo model based
on the Babcock-Leighton $\alpha$-effect, \cite{Choudhuri2004ApJ} found that at the start of cycles, helicities tend to be opposite of the preferred hemispheric trends. Temporary changes in the helicity hemisphere pattern have also been recorded in observations \citep{zhang2010}.

The interpretation of our results in terms of ``preferred'' states of linkage may also provide a new interpretation of the origin of the hemisphere rule of magnetic helicity. Our results also suggest that imbalances in this rule can be caused by changes in the linkage of large-scale field (i.e. larger than the small-scale field considered in mean-field models). More work is needed to develop this area, but the interpretation of helicity in terms of toroidal/poloidal linkage may help to develop existing mean-field models which attempt to explain hemisphere imbalances \citep[e.g.][]{yang_pipin_sokoloff_kuzanyan_zhang_2020}.

In future work, in order to perform a closer comparison with solar models and observations, we will move beyond Boussinesq MHD and study the global magnetic helicity in anelastic models. This will enable us not only to mimic solar parameters more closely, but also the parameters of other specific stars.

\section*{Acknowledgements}
Numerical computations were performed using the DiRAC Extreme Scaling service at the University of Edinburgh, operated by the Edinburgh Parallel Computing Centre on behalf of the STFC DiRAC HPC Facility (www.dirac.ac.uk). This equipment was funded by BEIS capital funding via STFC capital grant ST/R00238X/1 and STFC DiRAC Operations grant ST/R001006/1. DiRAC is part of the National e-Infrastructure.
PG acknowledges PhD sponsorship TRAW6PAA8 from the School of Mathematics and Statistics, the University of Glasgow.


\newcommand{\noopsort}[1]{} \newcommand{\printfirst}[2]{#1}
  \newcommand{\singleletter}[1]{#1} \newcommand{\switchargs}[2]{#2#1}

\appendix

\section{Appendix}

In order to verify the correct implementation of the helicity formula in the code, the following test was performed. In a magnetically closed spherical shell, linear force-free solutions exist where
\begin{equation}\label{force-free}
    \bnab\times\Bv=\lambda\Bv,
\end{equation}
with constant $\lambda$. The values of $\lambda$ that satisfy equation (\ref{force-free}) in the given domain form a discrete spectrum of eigenvalues for the curl operator. The corresponding eigenfunction of each eigenvalue has the form (in spherical coordinates)
\begin{subequations}
\begin{align}
    B_r =& \frac{1}{r^{3/2}}\big[c_1J_{3/2}(\lambda r)+c_2Y_{3/2}(\lambda  r)\big]\cos\theta, \\
    B_\theta =& -\frac{1}{2\lambda^{3/2}r^3}\left\{\sin(\pi-\lambda r)\left[c_1\left(\sqrt{\frac{2}{\pi}}\lambda^2 r^2- \sqrt{\frac{2}{\pi}}\right) + \sqrt{\frac{2}{\pi}}c_2\lambda r\right] \right. \nonumber\\
    &\left.+\cos(\pi-\lambda r)\left[c_2\left(\sqrt{\frac{2}{\pi}}\lambda^2 r^2- \sqrt{\frac{2}{\pi}}\right)-\sqrt{\frac{2}{\pi}}c_1\lambda r\right]\right\}\sin\theta,\\
    B_\phi =& \frac{\lambda}{2\sqrt{r}}\big[c_1J_{3/2}(\lambda r) + c_2Y_{3/2}(\lambda r)\big]\sin\theta,
\end{align}
\end{subequations}
where $J_{3/2}(x)$ and $Y_{3/2}(x)$ are Bessel functions of the first and second kind, respectively, and $c_1$ and $c_2$ are constants determined from the boundary conditions \citep{cantarella2000spectrum}. 

\begin{figure}[t]
    \centering
    \subfigure(a){\includegraphics[height=35mm]{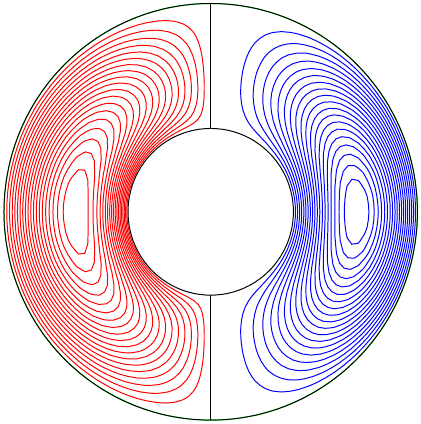}}
    \subfigure(b){\includegraphics[height=35mm]{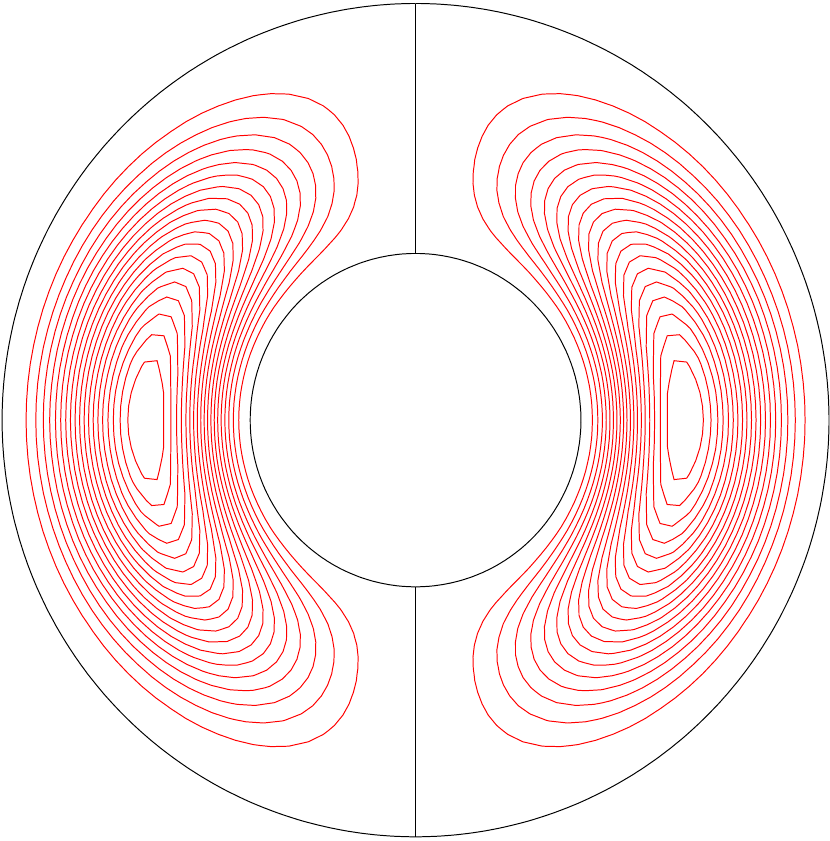}}
    \subfigure(c){\includegraphics[height=35mm]{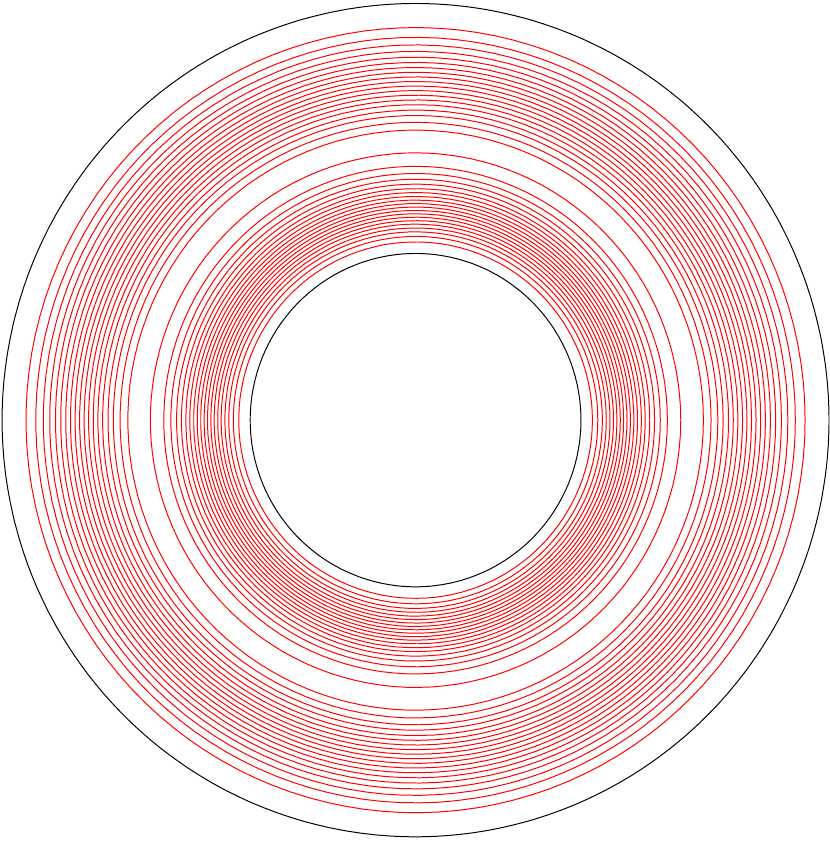}}
    \caption{(a) shows meridional lines of constant $\overline{B}_{\phi}$ (left half) and $r \sin \theta \dd_\theta \overline{h}$ (right half); (b) shows averaged helicity density (left half) and unaveraged helicity density (right half); (c) shows the helicity density in the equatorial plane. (Colour online)}
    \label{fig1}
\end{figure}

For a closed magnetic field, we have $B_r=0$ on the inner and outer boundaries of the spherical shell. The two conditions can be written as the matrix-vector system 
\begin{equation}
    \left[\begin{array}{cc}\displaystyle
    J_{3/2}(\lambda r_i)  & Y_{3/2}(\lambda r_i)  \\
     J_{3/2}(\lambda r_o) & Y_{3/2}(\lambda r_o) 
    \end{array}\right]\left[\begin{array}{c}
         c_1  \\
         c_2
    \end{array}\right]=\left[\begin{array}{c}
         0  \\
         0 
    \end{array}\right].
\end{equation}
For a non-trivial solution, we require that
\begin{equation}\label{eigen}
    J_{3/2}(\lambda r_i) Y_{3/2}(\lambda r_o) -Y_{3/2}(\lambda r_i) J_{3/2}(\lambda r_o)=0,
\end{equation}
and we consider the smallest values of $\lambda$ satisfying equation (\ref{eigen}). With $\lambda$ found, $c_1$ and $c_2$ are readily determined. For $r_i=2/3$ and $r_o=5/3$, $\lambda=3.383384$ and we take $c_1=1$ and $c_2=1.974996$.

Since this particular magnetic field is force-free, it satisfies $g=\lambda h$. Using this property, the magnetic helicity can be written as 
\begin{equation}\label{ff_hel}
    H= 2\lambda\int_V|\Lv h|^2\,\d V = \frac{2}{\lambda}\int_V|\Lv g|^2\,\d V.
\end{equation}
For this force-free field, the above relations represent an alternative way to calculate $H$ compared to the general equation (\ref{helicity}), and provide a useful test that the general formula has been coded correctly. For the values used in this work, we find, for the above force-free solution, that $H=6.91$ using either of equations \eqref{helicity} and \eqref{ff_hel}. 

In figure \ref{fig1}(b), the meridional plot of the azimuthal average of the helicity density is equal to that of a specific slice (shown on the right-had side). This is because the magnetic field is symmetric and, also, not dependent on $\phi$ in this example. However, this result is connected to a more general one regarding the magnetic helicity of any magnetic field in a spherical shell. That is,
\begin{equation}\label{az}
    \int_0^{2\pi}\Lv h\bdot\ev_\theta\,{\rm d}\phi = \int_0^{2\pi}\Lv g\bdot\ev_\theta\,{\rm d}\phi = 0,
\end{equation}
for given values of $r$ and $\theta$. The proof of this result can be most easily seen by expanding the integrands in terms of their spectral decompositions, e.g.
\[
\Lv h\bdot\ev_\theta = \sum^\infty_{l=0}\sum_{m=-l}^l \frac{{\mathrm i}m}{\sin\theta} H^m_l(r,t)P^m_l(\cos\theta){\mathrm e}^{{\mathrm i}m\phi}.
\]
The azimuthal average is found by setting $m=0$, hence confirming (\ref{az}). Thus, plots of the azimuthal average of the helicity density depend only on $(\Lv h\bdot\ev_\phi)(\Lv g \bdot \ev_\phi)$. This constraint is another way to check that magnetic helicity has been calculated correctly in the code. 


\end{document}